\documentclass{IEEEtran4PSCC}
\usepackage{cite}
\usepackage{float}
\usepackage{boldline} 
\usepackage{amsfonts}
\usepackage{makecell}
\usepackage{url}

\ifCLASSINFOpdf
   \usepackage[pdftex]{graphicx}
\else
\fi
\usepackage[cmex10]{amsmath}

\makeatletter
\let\old@ps@headings\ps@headings
\let\old@ps@IEEEtitlepagestyle\ps@IEEEtitlepagestyle
\def\psccfooter#1{%
    \def\ps@headings{%
        \old@ps@headings%
        \def\@oddfoot{\strut\hfill#1\hfill\strut}%
        \def\@evenfoot{\strut\hfill#1\hfill\strut}%
    }%
    \def\ps@IEEEtitlepagestyle{%
        \old@ps@IEEEtitlepagestyle%
        \def\@oddfoot{\strut\hfill#1\hfill\strut}%
        \def\@evenfoot{\strut\hfill#1\hfill\strut}%
    }%
    \ps@headings%
}
\makeatother

\begin{document}
\title{
Generating Multivariate Load States \\
Using a Conditional Variational Autoencoder
}

\author{
\IEEEauthorblockN{Chenguang Wang, Ensieh Sharifnia, Zhi Gao, Simon H. Tindemans, Peter Palensky}
\IEEEauthorblockA{Department of Electrical Sustainable Energy\\
Delft University of Technology\\
Delft, The Netherlands\\
\{c.wang-8, e.sharifnia, s.h.tindemans, p.palensky\}@tudelft.nl, z.gao-3@student.tudelft.nl}
}

\allowdisplaybreaks

\IEEEoverridecommandlockouts

\IEEEpubid{\parbox{\columnwidth}{\copyright 2022 IEEE. Personal use of this material is permitted. Permission from IEEE must be obtained for all other uses, in any current or future media, including reprinting/republishing this material for advertising or promotional purposes, creating new collective works, for resale or redistribution to servers or lists, or reuse of any copyrighted component of this work in other works.}\hspace{\columnsep}\makebox[\columnwidth]{ }}

\maketitle

\IEEEpubidadjcol

\begin{abstract}
For planning of power systems and for the calibration of operational tools, it is essential to analyse system performance in a large range of representative scenarios. When the available historical data is limited, generative models are a promising solution, but modelling high-dimensional dependencies is challenging. In this paper, a multivariate load state generating model on the basis of a conditional variational autoencoder (CVAE) neural network is proposed. Going beyond common CVAE implementations, the model includes stochastic variation of output samples under given latent vectors and co-optimizes the parameters for this output variability. It is shown that this improves statistical properties of the generated data. The quality of generated multivariate loads is evaluated using univariate and multivariate performance metrics. A generation adequacy case study on the European network is used to illustrate model's ability to generate realistic tail distributions. The experiments demonstrate that the proposed generator outperforms other data generating mechanisms. 
\end{abstract}


% Use this to place sponsorships
\thanksto{\noindent 
This work was supported by the Chinese Scholarship Council.
}

\section{Introduction}
In order to plan power systems and calibrate operational tools, it is essential to analyse system performance through a large range of representative scenarios \cite{bloomfield2021quantifying,panciatici2012operating}. 
Historical data is a key source of such scenarios, but when the available data set is too small for the desired application or when it cannot be made available for privacy reasons, it becomes valuable to have a model that can generate relevant data in abundant quantities. The challenge is that generated scenarios should embody both univariate distributions and multivariate inter-dependencies of the historical data \cite{Konstantelos2019}.

A common approach has been to fit parametric probabilistic models to historical scenarios, especially load states. In \cite{kang2007load}, \emph{Gaussian mixture models} (GMM) have been proposed to augment load data in distribution networks. Another study has introduced hidden Markov models to generate house-hold electric loads \cite{zia2011hidden}. More recently, a load generator has been designed using time-varying queuing models \cite{dos2020synthetic}. Due to the curse of dimensionality, it is especially challenging to use parametric methods for generation of high-dimensional states \cite[chapter 3]{curse_dimensionality}. Copula-based models are one class of generative models that does scale to higher dimensions, either using the Gaussian copula, or by `stacking' copulas in a vine structure, possibly in combination with dimension-reduction schemes \cite{Konstantelos2019}. 

As vine-based copula models are highly asymmetric and therefore prone to bias, it is appealing to investigate `native' high-dimensional models, such as neural networks. The \emph{variational autoencoder} (VAE) \cite{kingma2013auto} is an unsupervised machine learning model based on a deep neural network architecture. It has been successfully used in generating electricity load series, such as theft detection \cite{gong2020data} and electric vehicle load profiles  \cite{pan2019data}. However, the validation of generated data often remains limited to visual comparisons, which is not straightforward for  snapshots of larger and more complex electricity systems.

Moreover, most VAE implementations do not make full use \cite{yu2020tutorial} of the flexibility permitted by the mathematical framework in \cite{kingma2013auto}. Output noise tuning and training \cite{lin2019balancing,rybkin2021simple} has only recently been considered, with a focus on image and video data sets.
In power system related data generation applications, the output noise parameter is usually treated as a hyperparameter (i.e. a preset value that controls the learning process) \cite{doersch2016tutorial} and noise is not actually inserted into samples \cite{gong2020data, pan2019data, mylonas2021conditional,qi2020optimal,bregere2020simulating}.

This paper bridges those identified gaps by investigating the impact of output noise and its parameterisation, and by analysing generated data using performance metrics. This is done for the VAE and the \emph{conditional} VAE (CVAE), in the context of large-scale spatial load patterns of European countries. This lays the basis for wider applications of this method to synthetic load generation at lower aggregation levels, where consumption patterns are inherently more variable. 

The main contributions of this paper are:

\begin{itemize}
    \item We show how a sample-dependent output noise parameter can be co-optimised in the training process and how this noise is used in the generative process.
    \item  We put forward a set of data quality metrics for generative models, consisting of three statistical tests for univariate distributions and  multivariate dependencies.
    \item We introduce a simple multi-area adequacy assessment model that is used to test tail distributions. 
    \item Through comprehensive experiments, we show the performance and practicality of VAE- and CVAE-based load generators in comparison with \emph{Gaussian copula} and \emph{conditional generative adversarial network} (cGAN) models. 
\end{itemize}

\section{Data Generation Mechanism}{\label{sec:Mechanism}}

In this section, a representative multivariate load state generation mechanism is proposed, based on the \emph{conditional variational autoencoder} (CVAE).

\subsection{CVAE-based generative model}
The CVAE is a neural network architecture that is trained to learn the salient features of historical data by mapping (\emph{encoding}) historical system states onto a lower-dimensional latent space where the latent distribution is approximately normal - and transforming latent vectors back (\emph{decoding}) into a high-dimensional state space \cite{Kingma2019}. The decoder is used in conjunction with contextual information $c$ to generate representative states (which can be omitted to obtain a regular VAE model). Consequently, the model is able to generate samples with a similar distribution to the historical data, by transforming normally distributed samples in the latent space back to the data space. We note that the latent (i.e. hidden) representation of a data point is used solely to facilitate reconstruction and synthesis. It does not need to be imbued with a particular meaning.

\begin{figure} [t!p]
  \centering 
    \includegraphics[scale=0.55]{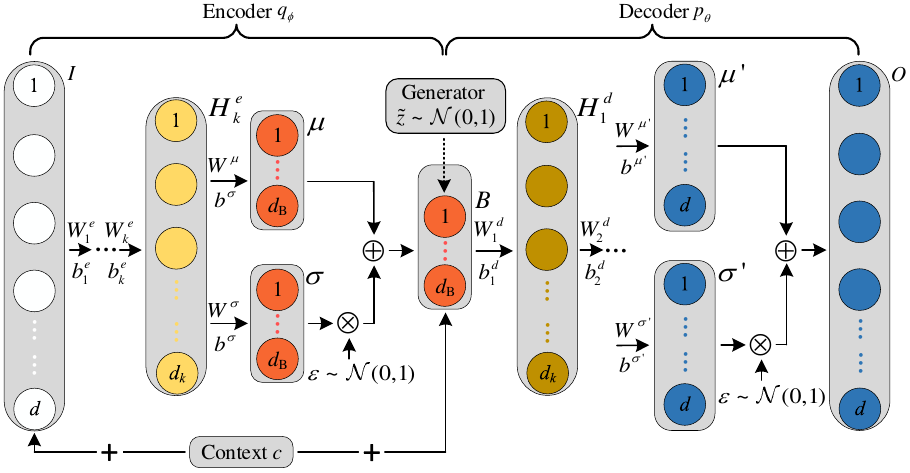}
  \caption{Schematic of the CVAE.}
  \label{fig:CVAE} 
\end{figure}

The structure of the CVAE algorithm is depicted in Fig.~\ref{fig:CVAE}. The \emph{encoder} maps the $d$-dimensional input data $x$ to the code $z$ in the lower-dimensional latent space through $k$ hidden layers $H_l^e$, $l=1,\ldots,k$. Weight matrices $W_{l}^{e}$, bias vectors $b_{l}^{e}$ and the context $c$ are utilized in the encoding process as
\begin{subequations}
		\begin{align}\label{eq:y}
		\left(\begin{matrix} \mu \\ \sigma \end{matrix}\right)  = & \left(\begin{matrix} 	
		W^{\mu} \\ W^{\sigma} \end{matrix}\right)(a(W_{k}^e(\ldots 	
		a(W_{1}^{e}(x,c) + b_{1}^{e})\ldots)+b_{k}^e)) \nonumber \\ &
		+ \left(\begin{matrix} b^{\mu} 	\\ b^{\sigma} \end{matrix}\right)\,,
		\\
	\label{eq:hatz}
	z =& \mu+ \epsilon \odot \sigma \, , 
	\end{align}
\end{subequations}
where $a$ represents an element-wise nonlinear activation function. Vectors $\mu$ and $\sigma$ parameterize an input-dependent normal distribution in the latent space. The output $z$ is sampled accordingly, using $\epsilon$, a vector that is sampled from a standard normal distribution, and the Hadamard product $\odot$. 

Mirroring the encoder network, the decoder maps the sampled latent space code $z$ to the $d$-dimensional output data $\hat{x}$ using
\begin{subequations}
	\begin{align}\label{eq:mu_2}
	\left(\begin{matrix} \mu' \\ \sigma' \end{matrix}\right) = & \left(\begin{matrix} W^{\mu'} \\ W^{\sigma'} \end{matrix}\right)(\ldots a(W_{1}^{d}(z,c) + b_{1}^{d})\ldots) + \left(\begin{matrix} b^{\mu'} \\ b^{\sigma'} \end{matrix}\right)\, ,
	\\
	\label{eq:hatx}
	\hat{x} =& \mu' + \epsilon \odot \sigma' \, , 
	\end{align}
\end{subequations}
where $W_{l}^{d}$ and $b_{l}^{d}$ denote weight matrices and bias vectors for decoding, respectively. $\mu'$ and $\sigma'$ parameterize a $z$-dependent normal distribution in the $x$ space.

\subsection{Training and generation process}

In the training stage, the whole structure of the CVAE model is utilized as Fig.~\ref{fig:CVAE}. Weight matrices $W$ and bias vectors $b$ are updated in an iterative way with the goal of minimizing the loss function \cite{Kingma2019}
\begin{align}\label{eq:loss_sum}
    \mathcal{L}=\mathcal{L}_{D_{KL}}+\mathcal{L}_{Re}.
\end{align}
The \emph{Kullback-Leibler loss} $\mathcal{L}_{D_{KL}}=\sum_i D_{KL}(q_\phi(z|x_i)||p(z))$ is the sum over all training data points $x_i$ (assumed i.i.d.) of the Kullback–Leibler divergence between that point's posterior distribution $q_\phi(z|x_i)$ and the prior distribution $p(z)$ (chosen as the standard normal distribution). The posterior distribution $q_\phi(z|x_i)$ is determined by the parameters $\phi$ of the encoder network and represents the mapping of the point $x_i$ into a normal distribution in the latent space using \eqref{eq:y} and \eqref{eq:hatz}. As the Kullback-Leibler divergence between two normal distributions can be evaluated directly \cite{doersch2016tutorial}, the Kullback-Leibler loss is computed as
\begin{align}\label{eq:KL_loss}
    \mathcal{L}_{D_{KL}}
    & = \frac{1}{2}\sum_{i=1}^n\sum_{j=1}^d(-1+{\sigma^2_{i,j}}+{\mu}^2_{i,j}-\log{\sigma^2_{i,j}}),%\nonumber
\end{align}
where $n$ denotes total number of observations used for training and $(\mu_i, \sigma_i)$ are evaluated for data point $x_i$ and condition $c$ using \eqref{eq:y}. 

The \emph{reconstruction loss} $\mathcal{L}_{Re}$ stands for the negative log-likelihood of reconstructing the inputs $x_i$ via their latent space codes and the decoder that is parameterized by $\theta$. The  reconstruction loss is thus computed as 
\begin{align}\label{eq:Re_loss}
    \mathcal{L}_{Re} & = - \sum_{i=1}^{n} \mathbb{E}_{Z\sim q_\phi(z|x_i)}[\log_{P_\theta}(x_i|Z)] \\ 
    & \approx \frac{1}{2}\sum_{i=1}^n\sum_{j=1}^d((x_{i,j}-\mu'_{i,j})^2/\sigma'^2_{i,j}+\log\sigma'^2_{i,j})+\frac{n d}{2}\log2\pi,\nonumber
\end{align} 
where the final step involves a single-point approximation of the expectation and $(\mu_i', \sigma_i')$ are obtained from the randomly generated latent code $z(x_i)$ and the condition $c$ using \eqref{eq:mu_2}. During training, the full-sample sum in loss functions \eqref{eq:KL_loss} and \eqref{eq:Re_loss} are replaced by batch-sample averages. The constant $\frac{n d}{2}\log2\pi$ of $\mathcal{L}_{Re}$ is omitted.

After the training process, only the decoder part of the trained CVAE network is utilized to generate data. Latent space codes $\Tilde{z}$ are sampled from the standard normal distribution $\mathcal{N}(0,I)$ (see Fig.~\ref{fig:CVAE}). Then, data space samples $\Tilde{x}$ are sampled from distribution $\mathcal{N}(\mu'(\Tilde{z},c),\sigma'(\Tilde{z},c))$, whose parameters are determined by $\Tilde{z}$ and $c$ using \eqref{eq:mu_2}. We note that although the amount of available training data determines the information contained within the model, there is no limit to the amount of data that can be generated.

In this way, a complex data distribution in the $x$ space is constructed as a continuous superposition of normal distributions that are parameterised by the normally distributed coordinate $z$. Using the procedure above, the encoder and decoder networks are trained to adapt any distribution to this normally distributed latent space. We note that other distributions besides the normal distribution can be used as the prior for the latent space coordinate $z$ \cite{xu-durrett-2018,joo2020dirichlet} -- selecting the best latent space representation for a particular class of problems remains an open research problem.

\subsection{Network and output noise co-optimization strategy}

It is common for CVAE implementations to generate data $\Tilde{x}$ not by sampling from $\mathcal{N}(\mu'(\Tilde{z},c),\sigma'(\Tilde{z},c))$ via \eqref{eq:hatx}, but by directly using the mean value $\mu'(\Tilde{z},c)$ (the maximum likelihood sample). Moreover, the standard deviation $\sigma'$ is not co-optimized in the training process of \eqref{eq:loss_sum}, but considered a hyperparameter that fixes $\sigma'_{i,j}$=$s$ identically in all dimensions, so that \eqref{eq:Re_loss} can be replaced by
\begin{align}\label{eq:Re_loss_revised}
    \Tilde{\mathcal{L}}_{Re} = \frac{1}{2}\sum_{i=1}^n\sum_{j=1}^d \frac{(x_{i,j}-\mu'_{i,j})^2}{s^2}.
\end{align}
In contrast, we investigate the model in which the parameters $\sigma'$ of the output noise distribution are co-optimized as a function of $z$ during training, as was recently also (independently) proposed in \cite{rybkin2021simple}. In addition, we explicitly add output noise $\epsilon \odot \sigma'(\Tilde{z},c)$ to the generated data. To compare the different approaches, the quality of the generated data is evaluated under all four combinations (Table~\ref{tab:strategies}): whether $\sigma'$ is co-optimized in the training stage (Auto $\sigma'$) or set to a fixed value (Fixed $\sigma'$); whether the noise $\epsilon \odot \sigma'(\Tilde{z},c)$ is added to the outputs (Noisy) or not (Noise free).
 
\subsection{Loss function weight tuning strategy} {\label{sec:beta}}

The two loss terms have opposing effects. The Kullback-Leibler loss $\mathcal{L}_{D_{KL}}$ ensures a good fit with the prior distribution that samples are generated from, thus suppressing spurious generated points at the expense of `smoothing' the output. The reconstruction loss $\mathcal{L}_{Re}$, on the other hand, promotes exact reconstruction of the training data. In this paper, in addition to the output noise, we also study the effect of a heuristic weighting factor $\beta$ \cite{burgess2018understanding} for the Kullback-Leibler loss term $\mathcal{L}_{D_{KL}}$ on statistical properties of the generated data.

All aforementioned combined strategies are explicated in Table~\ref{tab:strategies}. Their impacts  to the quality of generations will be investigated in the following sections. Particularly, the settings of standard deviation $\sigma'$ and weight $\beta$ influence the objective function in the training process and will ultimately affect the generated data. On the other hand, the use of output noise, $\epsilon \odot \sigma'(\Tilde{z},c)$, will directly impact the data generation stage. 

\renewcommand{\arraystretch}{1.7} %
\begin{table}[htp] 
\caption{Overview of model permutations used in experiments} 
\centering
\begin{tabular}{p{0.135\textwidth}p{0.125\textwidth}p{0.13\textwidth}}
                  \hlineB{4} %
  \textbf{Strategy (with $\beta$)}           & \textbf{Objective function}  & \textbf{Generation}  \\\hline
Auto $\sigma'$, Noisy         & $\beta\mathcal{L}_{D_{KL}} + \mathcal{L}_{Re}$          &  $\mathcal{N}(\mu'(\Tilde{z},c),\sigma'(\Tilde{z},c))$\\
Auto $\sigma'$, Noise free    & $\beta\mathcal{L}_{D_{KL}} + \mathcal{L}_{Re}$          &  $\mu'(\Tilde{z},c)$   \\
Fixed $\sigma'$,  Noisy       & $\beta\mathcal{L}_{D_{KL}} + \Tilde{\mathcal{L}}_{Re}$  &  $\mathcal{N}(\mu'(\Tilde{z},c),s I)$      \\
Fixed $\sigma'$,  Noise free  & $\beta\mathcal{L}_{D_{KL}} + \Tilde{\mathcal{L}}_{Re}$  &  $\mu'(\Tilde{z},c)$   \\ 
                \hlineB{4}
               \end{tabular}
               \label{tab:strategies}
\end{table}

\section{Case study: European load data}

In this section, the performance of our proposed CVAE-based generative model is analysed using a European load data set. This is done with three data quality metrics that measure univariate distributions and multivariate dependencies. Impacts to the quality of generated data are investigated under the experimental settings in Table~\ref{tab:strategies}, using both conditional and regular VAEs.

\subsection{Data source and generation}

Historical hourly load data for 32 European countries between 2013 and 2017 was obtained from the Open Power System Data platform \cite{Muehlenpfordt2019} (package version \emph{2019-06-05}). Columns of AL, CS, CY, GB, TR and UA were dropped for incomplete records. The historical data was randomly split into training and test set in blocks of one week with the proportion of 4:1 (35,148 training and 8,569 test samples). The training set was min-max normalized before being fed into the CVAE model and the inverse transformation was applied to generate samples. The context information $c$ is the hour of day. Both total and hourly volumes of the generated data are the same with the training data set, in order to balance them for visual and statistical analysis. However, we emphasise that the purpose of constructing such a generative model is to have the ability to generate limitless non-repeating data, e.g. for reducing the risk of overfitting in downstream machine learning tasks.

The parameters of the generative models were tuned for optimal performance, for both the VAE and the CVAE. The network contained 2 hidden layers in the encoder with dimensions of 24 and 16, respectively; the bottleneck layer had 8 nodes (8-dimensional latent vector). The decoder also had 2 hidden layers with the same dimensions in reverse order. Comparisons against 4-neuron and 16-neuron bottleneck revealed that a smaller bottleneck results in excessive loss, whereas a larger bottleneck insufficiently forces the network to learn features. In the CVAE model, the hourly time-of-day was encoded cyclically using sine/cosine representation. 
 
The ReLU activation function was used, except for the generation of $\mu$ and $\sigma$ leading up to the bottleneck and output layers. The \emph{adaptive moment estimation} (Adam) weight optimizer \cite{kingma2014adam} was utilized with default settings to iteratively optimize the value of weight matrices $W$ and bias vectors $b$. The batch size and learning rate related parameter $\alpha$ for training was 64 and $10^{-4}$ respectively and $20,000$ training iterations were used. Training and data generation of the model was conducted in Python using \texttt{tensorflow} on the Google Colab environment using the GPU option. The code used for this paper is available for download\cite{wang2022CVAE-code, MC-PSCC2022}. 

\subsection{Data quality metrics} \label{sec:metrics}

To test a generative model's ability to reproduce the features of historical data, especially in high dimensions, statistical tests are required. Three tests are put forward to examine different aspects of the generated data set, in comparison with the historical data.

\subsubsection{Kolmogorov-Smirnov  test  for  univariate  marginal  distributions} \label{sec:K-S test}

We used the two-sample Kolmogorov-Smirnov (K-S) test  \cite{massey1951kolmogorov}  to see whether the generated data was able to reproduce the marginal load distributions for each of the countries in the data set. For a given output dimension (load in a single country), historical and generated data are compared. Under the null hypothesis that historical and generated data are drawn from the same model, the $p$-values should follow a uniform distribution. In other words, when the historical data is compared against itself, the cumulative distribution of $p$-values should lie on the diagonal. Thus, for generative models, the closer the cumulative distribution of $p$-values lies to the diagonal, the higher the similarity between the two distributions. 

Clearly, the models are unlikely to exactly reproduce the historical distribution, thus large deviations from the ideal curve will show up for large-sample tests. Nevertheless, to analyse the degree of performance of various models, we use repeated tests on smaller sample sets that result in clear differentiation, as in \cite{Konstantelos2019}. In this paper, 0.5{\%} of the data set, i.e. 176 data points out of 35,148, were randomly drawn from training and generated data set, and then a $p$-value was calculated accordingly. This process was  repeated 5,000 times for each country and a curve was constructed from all $p$-values.

\subsubsection{Autoencoder-based point-wise test for multivariate dependencies}

Autoencoder (AE) neural networks have been proven to be highly sensitive anomaly detectors \cite{wang2020detection}. Unlike (C)VAE networks, AEs have no stochastic layers and only minimize the reconstruction loss $r= \sum_i \lVert {x_i}-{\hat{x_i}}\rVert^{2}/d$. An AE learns to compress and decompress the data based on properties of the training set. As a result, data points with dependencies that deviate substantially from that in the training set tend to have higher reconstruction errors.

A separate AE network was trained for this test, with hyperparameters equal to that of the CVAE model, except for the stochastic layers and objective function. Reconstruction errors of all data points (historical or generated) are plotted as cumulative distributions for easy comparison. 
As a test for overfitting of the autoencoder on the training data, the autoencoder test was performed on the training and test data. The two distributions visually overlapped, suggesting this is not a concern.

\subsubsection{Energy test for multivariate dependencies of population}

Another two-sample test, the energy test \cite{energy_test}, was conducted to examine whether the \emph{multivariate dependencies} of the population were well acquired from the training set. The energy test, computed using the PyTorch library torch-two-sample \cite{code_energytest} uses a user-specified number of permutations (200 was used) to calculate a $p$-value. The same as for the K-S test, we used random subsets of 0.5\% of the generated population and historical population. We repeated this process 1,000 times to draw a distribution of $p$-values and compare it with the uniform distribution (which would be expected if the data was drawn from the historical distribution). 

\begin{figure} [t!p]
  \centering 
    \includegraphics[scale=0.24]{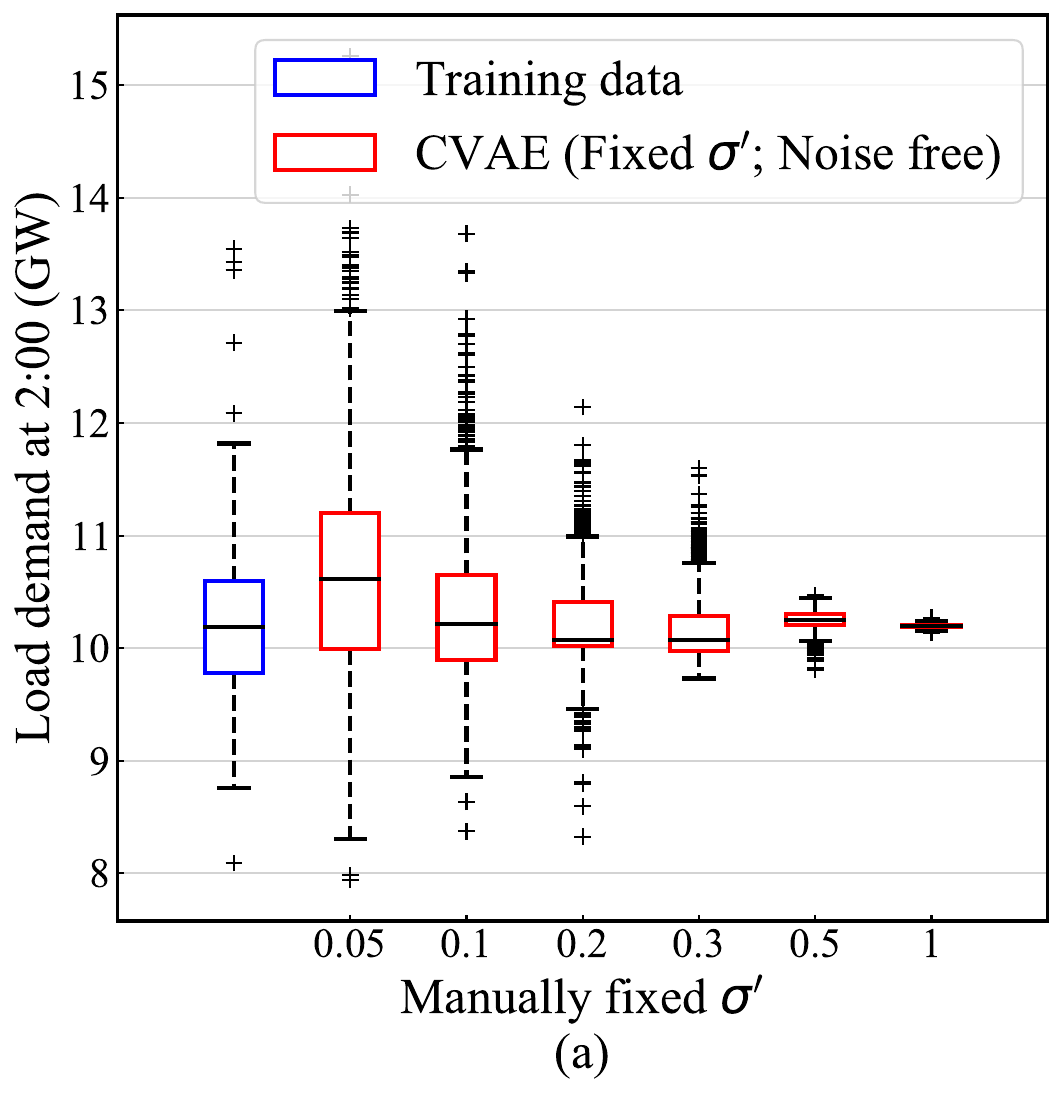}
    \includegraphics[scale=0.24]{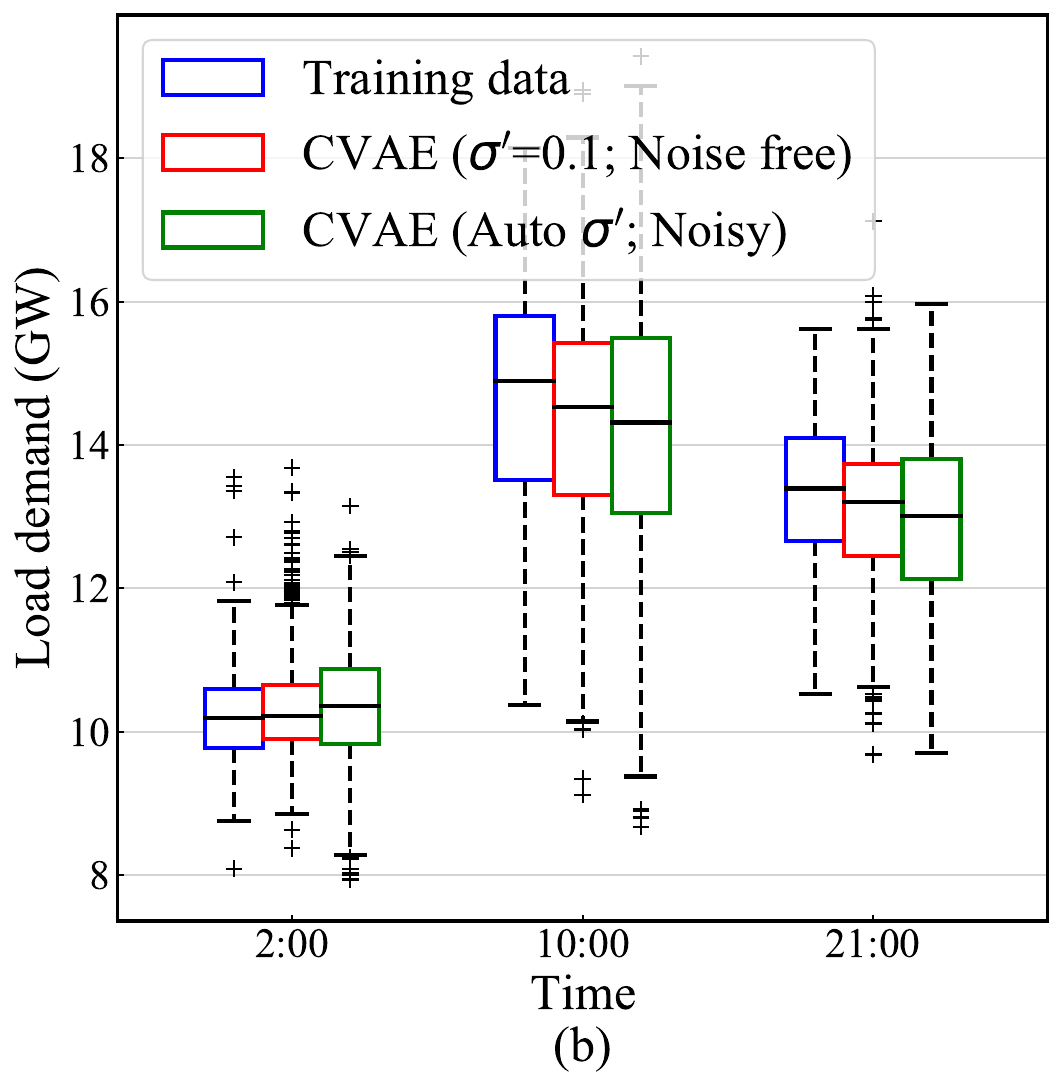}
  \caption{(a) Box plots of the original and generated load values in the Netherlands, based on 1459 data points at 2:00. Data was generated by CVAE using `Fixed $\sigma'$, Noise free' approach with different $\sigma'$. (b) Distribution comparison of the original and generated load data in the Netherlands, based on 1459 (2:00), 1465 (10:00) and 1465 (21:00) data points, respectively. Data was generated by CVAE using `Fixed $\sigma'$, Noise free' method ($\sigma'$=0.1) and `Auto $\sigma'$, Noisy' scheme.}
  \label{fig:Box-plot-Visual} 
\end{figure}

\begin{figure}  [t!p]
  \centering 
    \includegraphics[width=\linewidth]{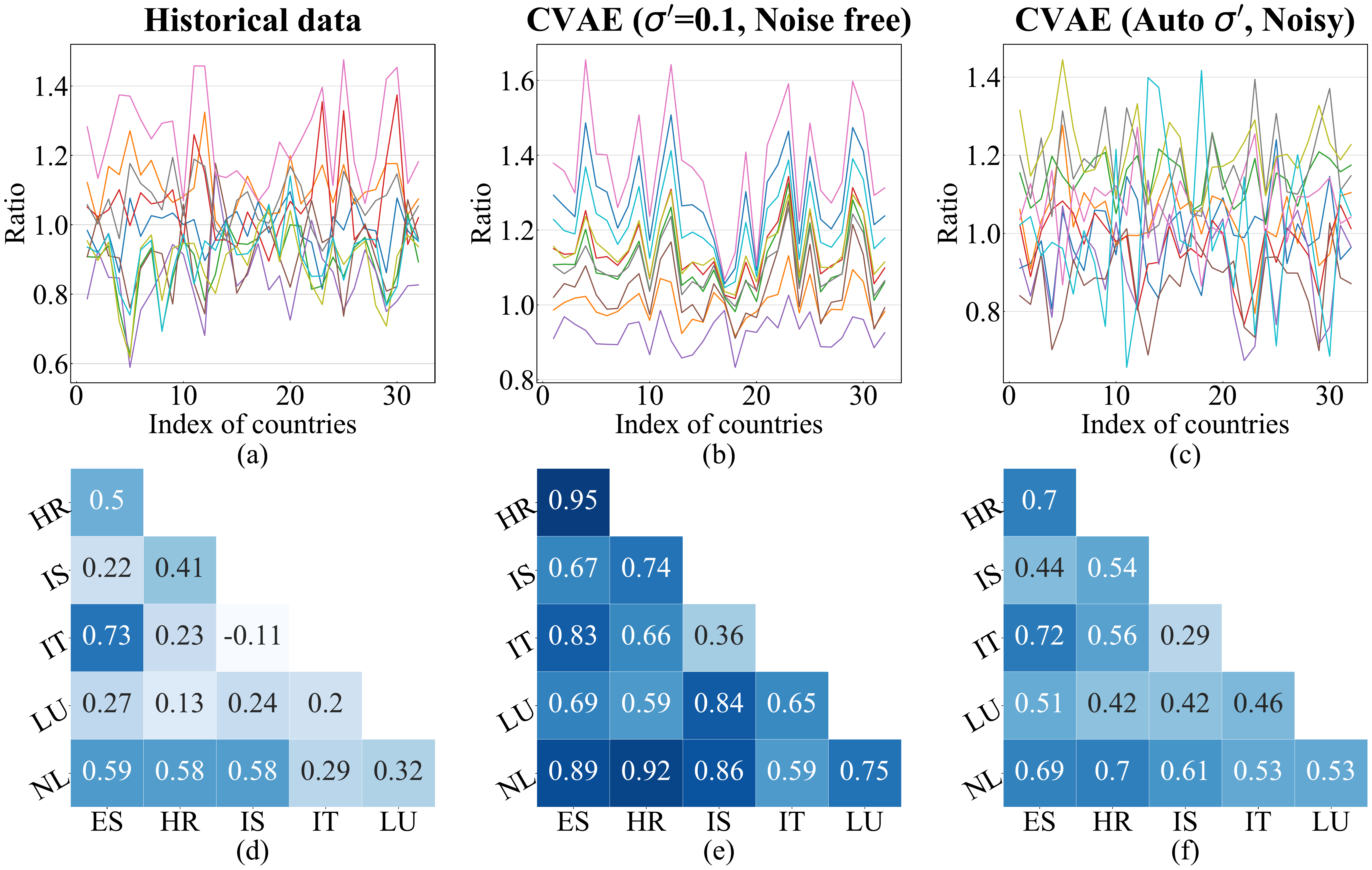} 
  \caption{(a), (b) and (c) display 10 typical ratios of 32 countries' historical and generated data to the historical mean values at 2:00. (d), (e) and (f) demonstrate the Pearson correlation coefficient matrices of 6 (out of 32) countries'  historical and generated data at 2:00. The horizontal and vertical dimensions in the matrices are Spain (ES), Croatia (HR), Iceland (IS), Italy (IT), Luxembourg (LU) and the Netherlands (NL). 
  }
  \label{fig:Correlation Mtrix1} 
\end{figure}

\begin{figure*}  [t!p]
    \centering 
    \includegraphics[width=\linewidth]{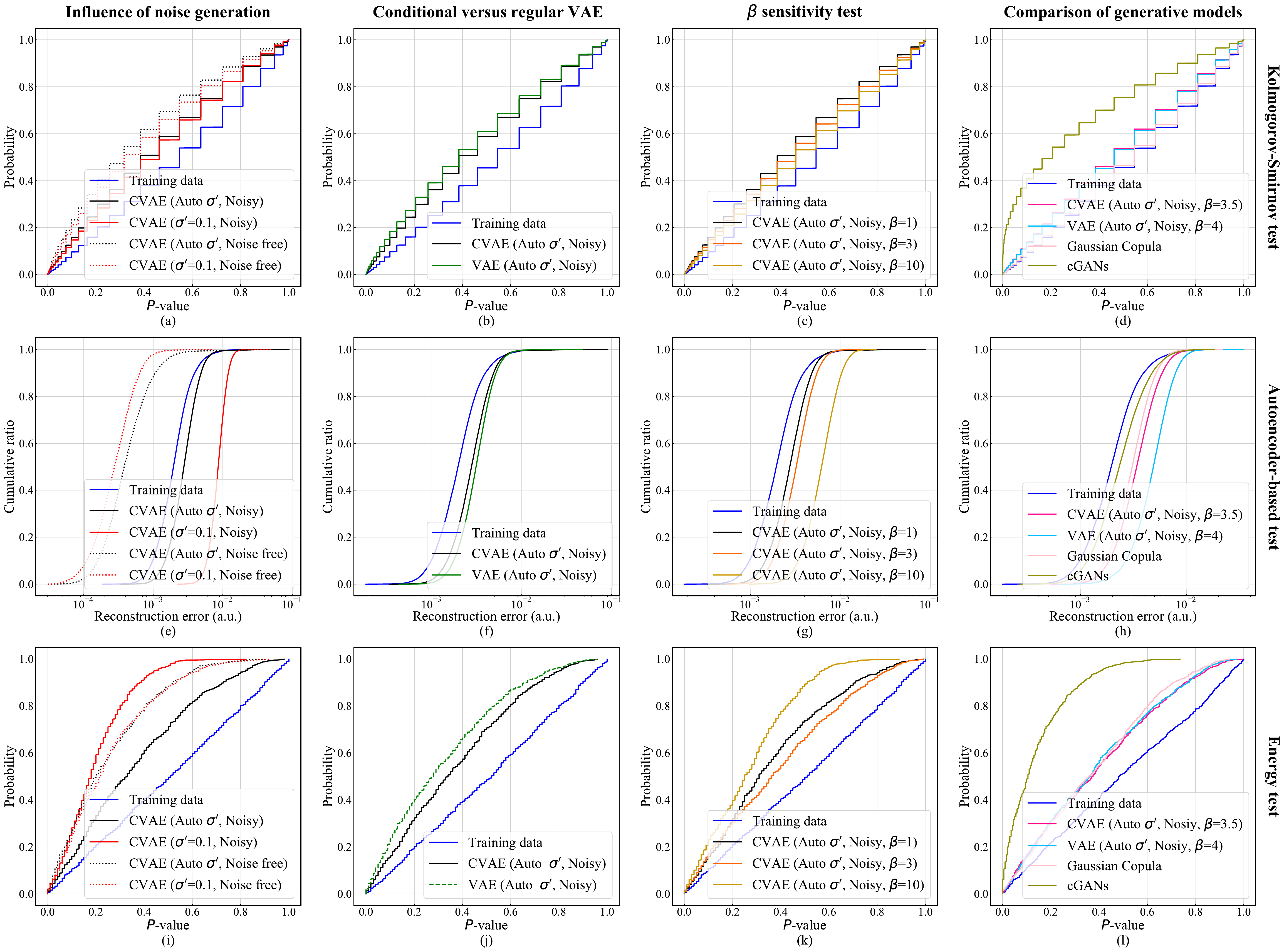}
    \caption{Results of statistical tests. Each column denotes a set of experiments (noise generation, training condition, value of $\beta$ and model family). The three rows depict results for the three tests described in Section~\ref{sec:metrics}.
    }
    \label{fig:Comparision}  
\end{figure*}

\subsection{Results}

\subsubsection{Visual comparison of univariate distributions}

In this experiment, the CVAE with fixed $\sigma'$ and no output noise was used to generate 1,459 load demands, conditioned on the time 2:00. Results for the Netherlands are shown in Fig.~\ref{fig:Box-plot-Visual}a, for various values of $\sigma'$. As the output noise assumed in training increases, the variability of the generated points decreases (because noise is not actually added). When $\sigma'$ = 0.1, the distribution of generations is the closest to that of historical data. This setting will be used for all further experiments with fixed $\sigma'$.\footnote{Note that we only fix a single parameter in this case. An approximate visual match of the box plots is a necessary condition for a good overall fit, justifying the choice $\sigma'$ = 0.1 for this comparison.} 

Fig.~\ref{fig:Box-plot-Visual}b further compares data generated using the `Fixed $\sigma'$' and `Auto $\sigma'$, Noisy' schemes and the training data. Conditioning on 2:00, 10:00 and 21:00 was performed, and results are shown for the Netherlands. Both methods are able to qualitatively reproduce the features of the data.

\subsubsection{Multivariate correlations}

The top row of Fig.~\ref{fig:Correlation Mtrix1} shows the loads of all countries for 10 different snapshots at 2:00, relative to the mean load in those countries at 2:00.  
Compared to historical data (a) and the noisy generator (c), samples generated by the noise-free generator clearly show higher correlations between countries. This is confirmed by the correlation analysis between six countries in the bottom row of Fig.~\ref{fig:Correlation Mtrix1}. 
By omitting output noise, the noise-free generator generated (too) highly correlated samples. 

Sensitive experiments for the multivariate dependencies will be conducted in the following sections using the autoencoder-based point-wise test. The accurate representation of multivariate dependencies will be important for the analysis of supply shortfalls in Section~\ref{sec:adequacy}.

\subsubsection{Influence of noise generation}

In this experiment, the influence of the four strategies listed in Table~\ref{tab:strategies} were tested with $\beta=1$. Results for the statistical tests described in Section~\ref{sec:metrics} are shown in the first column of Fig.~\ref{fig:Comparision}. The K-S test results and autoencoder results show that the inclusion of output noise is essential to improve marginal distributions (Fig.~\ref{fig:Comparision}a) and increase output variability to the level of the historical data (Fig.~\ref{fig:Comparision}e). 
In addition, the autoencoder and energy tests show that automatic tuning of the noise strength (Auto $\sigma'$) is essential to improve the multivariate dependencies of the generated samples. Together, this experiment shows that the `Auto $\sigma'$, Noisy') generator outperforms the other approaches listed in Table~\ref{tab:strategies}. This was to be expected given the mathematical theory behind the CVAE (which includes noise), but is at odds with common implementations. 

\subsubsection{Comparison between conditional and regular VAE}

In the second column of Fig.~\ref{fig:Comparision}, the performance of the CVAE and VAE models (with $\beta=1$) is compared. The CVAE model slightly outperforms the VAE model in all categories. One possible explanation is that the CVAE model has access to the context $c$ (time of day), which effectively increases the dimension of the latent space. Because of its better performance, we continue using the CVAE model in subsequent experiments, but the results suggest that a VAE model delivers comparable performance, and may be preferable when no natural conditioning variable is available. 

\subsubsection{$\beta$ sensitivity test}

The third column of Fig.~\ref{fig:Comparision} shows the impact of $\beta$ (values 1, 3, 10) on the performance of the CVAE (Auto $\sigma'$, Noisy) model. As $\beta$ is increased, the performance on the K-S test (Fig.~\ref{fig:Comparision}c) improves, indicating an improved ability to learn  marginal distributions. On the other hand, performance on the autoencoder test (Fig.~\ref{fig:Comparision}g) worsens, suggesting that points `outside' of the training point cloud are generated for large $\beta$. Finally, the energy test (Fig.~\ref{fig:Comparision}k) indicates that a moderate value of $\beta$ can strike a balance between the opposing requirements: the curve for $\beta=3$ is closest to the desired result. 
Nevertheless, depending on the application, it may be desirable to choose $\beta$ larger or smaller. 

\subsubsection{Comparison of Generative Models}

In the fourth column of Fig.~\ref{fig:Comparision}, the quality of data sampled from different generative models was investigated. The values of $\beta$ for CVAE and VAE models (both Auto $\sigma'$, Noisy) were tuned for optimal performance on the energy test (see previous section). In addition, \emph{Gaussian copula} \cite{liu2012high} and cGAN \cite{goodfellow2014generative} models were included for comparison.
The basic cGAN model was modified to use Wasserstein losses\cite{arjovsky2017wasserstein}. Both its generator and discriminator are deep neural networks; each has two hidden layers of 256 neurons, all activated with LeakyReLU ($\alpha$ = 0.2) except in the output layers, where linear and sigmoid activation functions are used for the generator and discriminator. Weights of the neurons are optimized with \emph{root mean square
propagation} (RMSprop) available from python package Keras. 

The K-S test shows the outstanding ability of the Gaussian copula model to reproduce marginal load distributions (a design feature of copula models \cite{Konstantelos2019}). This model also shows competitive performance on the autoencoder and energy tests. However, it will become clear in Section~\ref{sec:adequacy} that its tail-performance is worse than that of the (C)VAE models. The cGAN model shows the best performance on the autoencoder test, indicative of its ability to generate samples with realistic features. However, the model significantly underperforms on the K-S and energy tests, which suggests that the generated samples, though 'realistic', are unevenly distributed through the space of possible states. The optimised CVAE and VAE models are competitive on all three tests, with the CVAE model slightly outperforming the VAE model.

%%-------------------------------
\section{multi-area adequacy assessment} \label{sec:adequacy}

Next, we investigated the performance of the load generation mechanisms by using it for a multi-area adequacy assessment study, based on the ENTSO-E Mid-term Adequacy Forecast 2020 (MAF2020) \cite{ENTSO2020}. 
Multi-area adequacy assessment measures the sufficiency of generating capacity compared with the load on each of the nodes in the power system under transmission constraints. This can be considered a stress test of the generative model, as the outcomes are sensitive to high-load events (tail distributions) \emph{and} their dependencies between countries. 

Monte Carlo simulations were used to estimate \emph{Loss Of Load Expectation} (LOLE [h/year]) and \emph{Expected Energy Not Served} (EENS [MWh/year]). LOLE is the expected number of hours per year during which the supply does not meet demand. EENS is the expected amount of energy demand per year that cannot be supplied. Parameters from the MAF2020 study were used to construct a model for generating capacity and net transfer capacities between countries. They were combined with generated and historical load data to define a probabilistic model for the Monte Carlo simulations. We emphasise that the model thus constructed is not meant to be an accurate representation of the European grid, but a stylised problem that serves as a comparative testing ground for the generative models.

\subsection{Multi-area adequacy assessment structure}

We consider the network as a directed graph (to allow for asymmetric flow limits) where nodes are zones, edges are connections between zones, and edge capacities are transfer capacities. Each sampled state $w$ is represented by the available generating capacity $\overline{g}_{i}^w$ and demand $d_{i}^w$ of each node $i$. Based on the flow constraints and dispatching policy, the consumed power $p_{i}^w$ and load curtailment $c_{i}^w$ for each node can be calculated, related by
\begin{equation}
\label{eqn:loadCurtailment}
c_{i}^w = \max(0, d_{i}^w - p_{i}^w).
\end{equation}
We determine $c_i^w$ (and implicitly $p_i^w$) by solving a quadratic problem with variables $\tilde{c}_i$ (curtailment) and $\tilde{f}_{ij}$ (flows), which aims to minimize the total load curtailments and assumes that curtailments are balanced between areas \cite{evans2020assessing}, relative to the demand in that area:
\begin{align}
\label{eqn:objectiveFunction}
\vec{c}^w =\underset{\tilde{f}, \tilde{c}}{\text{arg~min}}  \sum_{i\in \mathcal{N}} \frac{1}{2d_{i}^w}\tilde{c}_{i}^2 +  \tilde{c}_{i} \\
\label{eqn:consa}
\underline{f_{ij}} \leq {\tilde{f}_{ij}} \leq \overline{f_{ij}}, & &\forall (ij) \in \mathcal{L} \\
\label{eqn:consb}
0 \leq \tilde{c}_{i} \leq d_{i}^w, &&\forall i \in \mathcal{N} \\
\label{eqn:consc}
d_{i}^w-\overline{g}_{i}^w \leq \sum_{j < i} \tilde{f}_{ji}-\sum_{j > i} \tilde{f}_{ij} + \tilde{c}_{i} \leq d_{i}^w ,&& \forall i \in \mathcal{N}
\end{align}
%Where $\alpha, \beta$ are positive constants. 
Here, $\mathcal{L}$ and $\mathcal{N}$ are the sets of connections (from $i$ to $j$ with $i<j$) and areas respectively. Constraints on power flow $\tilde{f}_{ij}$ from node $i$ to node $j$ are given in \eqref{eqn:consa}; \eqref{eqn:consb} limits curtailment and \eqref{eqn:consc} enforces flow and generating power constraints. The objective function (\ref{eqn:objectiveFunction}) has positive definite structure and the constraints are linear, so this optimization problem is strictly convex and has a unique solution. This optimization problem was solved using the python package quadprog \cite{pypi}.

\begin{table*}[t!]{\arraystretch}{1.0}
\centering
    \label{tab:resource_adequacy}
    \caption{Calculated risks of selected countries (with and without interconnection) using historical data and all generative models.}
    \begin{tabular}{p{1.45cm} p{1.1cm} p{0.9cm} p{0.9cm} p{0.9cm} p{0.9cm} p{0.2cm}  p{1.3cm} p{1.3cm} p{1.2cm} p{1.3cm} p{1.3cm}}
    
    \hlineB{4}
    Country&\multicolumn{5}{l}{LOLE (h/y)}& & \multicolumn{5}{l}{EENS (MWh/y)}\\
    \cline{2-6}  \cline{8-12}
    & Historical data & CVAE& VAE& Gaussian copula& cGAN& & Historical data & CVAE& VAE& Gaussian copula& cGAN\\
    \hline
    AT& 0.03(2)&0(0)&0.18(4)&0.32(5)&0.05(2)& &4(2)& 0(0)&33(11)& 42(10)& 10(5) \\
    NL& 0.74(8)&0.12(3)&0.80(8)&4.6(2)&3.6(2)& &119(17)&33(11)&300(45)&1135(67)& 1743(129) \\ 
    UK&37.8(6)&50.6(7)&54.1(7)&50.2(7)&223.6(14)& &5.20(10)$\cdot 10^4$ & 8.77(15)$\cdot 10^4$ &1.16(2)$\cdot 10^5$& 7.75(14)$\cdot 10^4$ & 4.99(4)$\cdot 10^5$\\
    \hline
    AT (island)& 0.74(8)&1.01(9)&0.88(9)&0.80(8)&0.61(7)& &221(33) &435(54)&334(45)& 273(40) & 255(41)\\
    NL (island)& 63.8(7)&65.2(8)&69.2(8)&69.4(8)&99.7(9)& &4.13(7)$\cdot 10^4$&4.35(7)$\cdot 10^4$&4.73(7)$\cdot 10^4$&4.61(7)$\cdot 10^4$& 6.74(8)$\cdot 10^4$\\
    UK (island)& 1026(3)&982(3)&884(3)&1033(3)&1965(4)& &4.28(2)$\cdot 10^6$&4.24(2)$\cdot 10^6$& 3.88(2)$\cdot 10^6$& 4.38(2)$\cdot 10^6$& 1.078(3)$\cdot 10^7$ \\
    \hlineB{4}
    \end{tabular}
    \label{tab:MLMC_speedup}
\end{table*}

\subsection{Power system model}

A European  adequacy assessment model was developed, based on the target year 2025 data from the ENTSO-E MAF2020 \cite{ENTSO2020}. The net transfer capacities between countries are defined as the summation of transfer capacities between their constituent zones, as reported in the MAF2020.
Since details of generators and unit capacities are not reported in the released dataset, we model the total generating capacity and the unit capacities in each country as follows. The assumed generating capacity of each country is a summation of conventional generating capacity in its zone(s) plus 5\% of nameplate wind power capacity. Unit sizes are set per country as the closest value under 500MW that is a divisor of the generating capacity; a unit availability of 83\% is used. Cyprus has no connection to other countries, so a unit capacity of 95MW is used to avoid excessive outages.

\begin{figure}[t!p]
	\centering
	\includegraphics[scale=0.345]{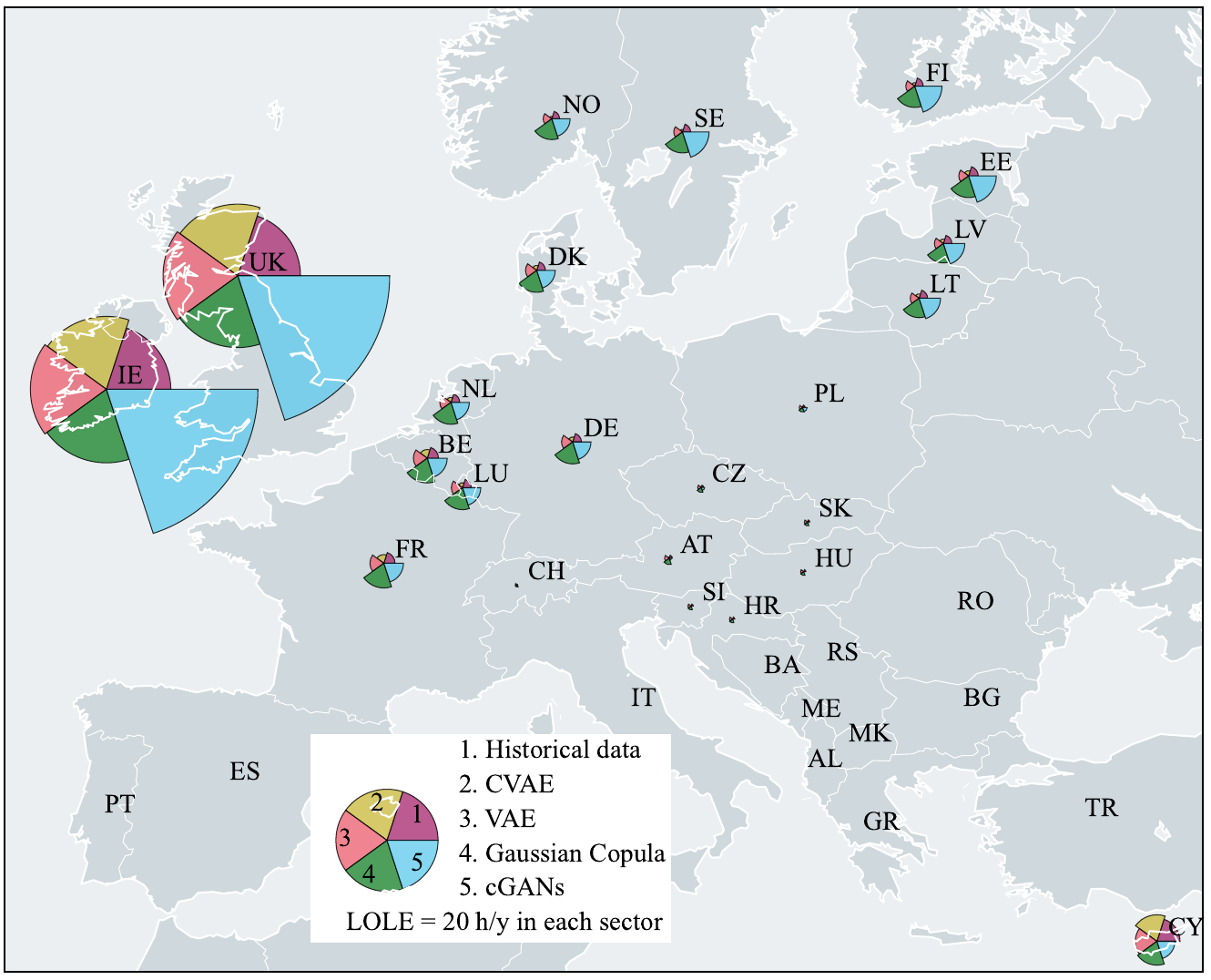}
	\caption{Comparison of LOLE estimates using historical load data and  generative load. The area of each sector of the disc represents the LOLE of the corresponding model (20h/y shown for scale in legend).}
	\label{fig:LOLE}
\end{figure}

\subsection{Multi-area adequacy assessment results}

To compare the CVAE, VAE, Gaussian Copula, and cGAN generators, they were trained on historical load data from 2017 and 2018 for 35 countries, retrieved from the Open Power System Data Platform (\cite{Muehlenpfordt2019}; columns for CS, IS and UA were omitted). Each model was used to generate 100,000 random load samples.
The `Auto $\sigma'$, Noisy' setting was utilized for the CVAE and VAE models, and $\beta$ was set to 10 for improved reproduction of the marginal distributions. 
For each model, 1,000,000 Monte Carlo generation samples were drawn and combined with random load samples to estimate the LOLE in each country. Fig.~\ref{fig:LOLE} depicts the estimated LOLE values for all generative methods and historical data. The area of each sector of the disc represents the LOLE obtained using a particular load states generating model. Numerical results for three countries with low (Austria, AT), medium (The Netherlands, NL) and high (UK) risk levels are shown in  Table~\ref{tab:resource_adequacy}. Standard errors for the least significant digits are shown in parentheses. Moreover, to investigate the beneficial effect of interconnection -- and therefore the importance of accurate multivariate modelling -- risks for these countries are also reported in the absence of interconnection (`island').   

By design, the Gaussian copula reproduces the marginal distributions of the historical data. Therefore, the calculated risk for \emph{islanded} systems is consistent with those for historical data. However, the results demonstrate that this model tends to overestimate risks for interconnected nodes (countries). 
The cGAN generative model tends to cause an overestimation of risks with \emph{both} islanded and interconnected nodes, sometimes very significantly (e.g.~the LOLE values for the UK and Ireland). In comparison, both the VAE and CVAE models generate data that results in risk estimates that are closer to those observed using historical data, although deviations exist from country to country. This suggests both models are able to substantially represent the multivariate tail distribution of the historical data.

The capability of generating load conditioned on hours is an additional advantage of CVAE in comparison with VAE in the adequacy assessment context. Load curtailments usually accrue during high load hours. So, time of day could be used as a control variable for an importance sampling Monte Carlo scheme that  preferentially samples load states at high load hours and compensates for the resulting bias by sample re-weighting. 

\section{Conclusions and future work}
In this paper, we have investigated the performance of CVAE- and VAE-based models to generate multivariate load states. Our inclusion of (1) sample noise in the generator and (2) co-optimised output noise parameters results in generated samples that show better marginal distributions and dependencies, when compared with common CVAE implementations (fixed noise parameter, noise omitted from the generator). A loss weighting factor $\beta$ (hyperparameter) can be used to tune the performance of the model. Performance was tested using three statistical tests and in a Monte Carlo generation adequacy study. The (C)VAE based models significantly outperformed Gaussian copula and cGAN models on at least one of these tests and were competitive on all others. 

With access to contextual information, the CVAE model slightly outperformed the VAE model. Moreover, such information can be used for targeted analysis, e.g. as part of a Monte Carlo importance sampling scheme that selects specific hours of the day. 

In future work, we will further investigate the universality of our proposed load generation scheme by applying it to lower load aggregation levels, such as household-level data. At this level, the privacy-preserving nature of synthetic data becomes very beneficial and should be carefully tested in addition to the distributional aspects. 

\section*{Acknowledgements}
We thank Kutay B\"olat for insightful discussions and the anonymous reviewers for helpful suggestions.

\bibliographystyle{IEEEtran}
\bibliography{literature}

\begin{thebibliography}{10}
\providecommand{\url}[1]{#1}
\csname url@samestyle\endcsname
\providecommand{\newblock}{\relax}
\providecommand{\bibinfo}[2]{#2}
\providecommand{\BIBentrySTDinterwordspacing}{\spaceskip=0pt\relax}
\providecommand{\BIBentryALTinterwordstretchfactor}{4}
\providecommand{\BIBentryALTinterwordspacing}{\spaceskip=\fontdimen2\font plus
\BIBentryALTinterwordstretchfactor\fontdimen3\font minus
  \fontdimen4\font\relax}
\providecommand{\BIBforeignlanguage}[2]{{%
\expandafter\ifx\csname l@#1\endcsname\relax
\typeout{** WARNING: IEEEtran.bst: No hyphenation pattern has been}%
\typeout{** loaded for the language `#1'. Using the pattern for}%
\typeout{** the default language instead.}%
\else
\language=\csname l@#1\endcsname
\fi
#2}}
\providecommand{\BIBdecl}{\relax}
\BIBdecl

\bibitem{bloomfield2021quantifying}
H.~Bloomfield, D.~Brayshaw, A.~Troccoli, C.~Goodess, M.~De~Felice, L.~Dubus,
  P.~Bett, and Y.-M. Saint-Drenan, ``Quantifying the sensitivity of {E}uropean
  power systems to energy scenarios and climate change projections,''
  \emph{Renewable Energy}, vol. 164, pp. 1062--1075, 2021.

\bibitem{panciatici2012operating}
P.~Panciatici, G.~Bareux, and L.~Wehenkel, ``Operating in the fog: Security
  management under uncertainty,'' \emph{IEEE Power and Energy Magazine},
  vol.~10, no.~5, pp. 40--49, 2012.

\bibitem{Konstantelos2019}
I.~Konstantelos, M.~Sun, S.~H. Tindemans, S.~Issad, P.~Panciatici, and
  G.~Strbac, ``Using vine copulas to generate representative system states for
  machine learning,'' \emph{IEEE Transactions on Power Systems}, vol.~34,
  no.~1, pp. 225--235, 2019.

\bibitem{kang2007load}
M.-S. Kang, C.-S. Chen, Y.-L. Ke, C.-H. Lin, and C.-W. Huang, ``Load profile
  synthesis and wind-power-generation prediction for an isolated power
  system,'' \emph{IEEE Transactions on Industry Applications}, vol.~43, no.~6,
  pp. 1459--1464, 2007.

\bibitem{zia2011hidden}
T.~Zia, D.~Bruckner, and A.~Zaidi, ``A hidden {M}arkov model based procedure
  for identifying household electric loads,'' in \emph{IECON 2011-37th Annual
  Conference of the IEEE Industrial Electronics Society}.\hskip 1em plus 0.5em
  minus 0.4em\relax IEEE, 2011, pp. 3218--3223.

\bibitem{dos2020synthetic}
F.~B. dos Reis, R.~Tonkoski, and T.~M. Hansen, ``Synthetic residential load
  models for smart city energy management simulations,'' \emph{IET Smart Grid},
  vol.~3, no.~3, pp. 342--354, 2020.

\bibitem{curse_dimensionality}
S.~Theodoridis, \emph{Machine Learning: A Bayesian and Optimization
  Perspective}, 2nd~ed.\hskip 1em plus 0.5em minus 0.4em\relax Academic Press,
  2020, {ISBN:978-0-12-818803-3}.

\bibitem{kingma2013auto}
D.~P. Kingma and M.~Welling, ``Auto-encoding variational bayes,'' \emph{arXiv
  preprint arXiv:1312.6114}, 2013.

\bibitem{gong2020data}
X.~Gong, B.~Tang, R.~Zhu, W.~Liao, and L.~Song, ``Data augmentation for
  electricity theft detection using conditional variational auto-encoder,''
  \emph{Energies}, vol.~13, no.~17, p. 4291, 2020.

\bibitem{pan2019data}
Z.~Pan, J.~Wang, W.~Liao, H.~Chen, D.~Yuan, W.~Zhu, X.~Fang, and Z.~Zhu,
  ``Data-driven {EV} load profiles generation using a variational
  auto-encoder,'' \emph{Energies}, vol.~12, no.~5, p. 849, 2019.

\bibitem{yu2020tutorial}
R.~Yu, ``A tutorial on {VAE}s: From {B}ayes' rule to lossless compression,''
  \emph{arXiv preprint arXiv:2006.10273}, 2020.

\bibitem{lin2019balancing}
S.~Lin, S.~Roberts, N.~Trigoni, and R.~Clark, ``Balancing reconstruction
  quality and regularisation in elbo for vaes,'' \emph{arXiv preprint
  arXiv:1909.03765}, 2019.

\bibitem{rybkin2021simple}
O.~Rybkin, K.~Daniilidis, and S.~Levine, ``Simple and effective {VAE} training
  with calibrated decoders,'' in \emph{International Conference on Machine
  Learning}.\hskip 1em plus 0.5em minus 0.4em\relax PMLR, 2021, pp. 9179--9189.

\bibitem{doersch2016tutorial}
C.~Doersch, ``Tutorial on variational autoencoders,'' \emph{arXiv preprint
  arXiv:1606.05908}, 2016.

\bibitem{mylonas2021conditional}
C.~Mylonas, I.~Abdallah, and E.~Chatzi, ``Conditional variational autoencoders
  for probabilistic wind turbine blade fatigue estimation using supervisory,
  control, and data acquisition data,'' \emph{Wind Energy}, vol.~24, pp.
  1122--1139, 2021.

\bibitem{qi2020optimal}
Y.~Qi, W.~Hu, Y.~Dong, Y.~Fan, L.~Dong, and M.~Xiao, ``Optimal configuration of
  concentrating solar power in multienergy power systems with an improved
  variational autoencoder,'' \emph{Applied Energy}, vol. 274, p. 115124, 2020.

\bibitem{bregere2020simulating}
M.~Br{\'e}g{\`e}re and R.~J. Bessa, ``Simulating tariff impact in electrical
  energy consumption profiles with conditional variational autoencoders,''
  \emph{IEEE Access}, vol.~8, pp. 131\,949--131\,966, 2020.

\bibitem{Kingma2019}
D.~P. Kingma and M.~Welling, ``{An Introduction to Variational Autoencoders},''
  \emph{Foundations and Trends in Machine Learning}, vol.~12, no.~4, pp.
  307--392, jun 2019.

\bibitem{xu-durrett-2018}
\BIBentryALTinterwordspacing
J.~Xu and G.~Durrett, ``Spherical latent spaces for stable variational
  autoencoders,'' in \emph{Proceedings of the 2018 Conference on Empirical
  Methods in Natural Language Processing}.\hskip 1em plus 0.5em minus
  0.4em\relax Brussels, Belgium: Association for Computational Linguistics,
  Oct.-Nov. 2018, pp. 4503--4513. [Online]. Available:
  \url{https://aclanthology.org/D18-1480}
\BIBentrySTDinterwordspacing

\bibitem{joo2020dirichlet}
W.~Joo, W.~Lee, S.~Park, and I.-C. Moon, ``Dirichlet variational autoencoder,''
  \emph{Pattern Recognition}, vol. 107, p. 107514, 2020.

\bibitem{burgess2018understanding}
C.~P. Burgess, I.~Higgins, A.~Pal, L.~Matthey, N.~Watters, G.~Desjardins, and
  A.~Lerchner, ``Understanding disentangling in $\beta$-{VAE},'' \emph{arXiv
  preprint arXiv:1804.03599}, 2018.

\bibitem{Muehlenpfordt2019}
\BIBentryALTinterwordspacing
J.~Muehlenpfordt, ``Time series,'' \emph{Open Power System Data}, 2019.
  Accessed on: Oct. 3, 2021. [Online]. Available:
  \url{https://data.open-power-system-data.org/time_series/2019-06-05}
\BIBentrySTDinterwordspacing

\bibitem{kingma2014adam}
D.~P. Kingma and J.~Ba, ``Adam: A method for stochastic optimization,''
  \emph{arXiv preprint arXiv:1412.6980}, 2014.

\bibitem{wang2022CVAE-code}
\BIBentryALTinterwordspacing
C.~Wang, ``Generating multivariate load states using a (conditional)
  variational autoencoder,'' 2022. Accessed on: Apr. 13, 2022. [Online].
  Available: \url{https://github.com/ChenguangWang-Sam/PSCC2022-CVAE}
\BIBentrySTDinterwordspacing

\bibitem{MC-PSCC2022}
\BIBentryALTinterwordspacing
E.~Sharifnia, ``Code release: System adequacy case study for {CVAE} load
  generation, {PSCC} 2022,'' 2022. Accessed on: Apr. 15, 2022. [Online].
  Available: \url{https://github.com/ensieh-sharifnia/MC-PSCC2022}
\BIBentrySTDinterwordspacing

\bibitem{massey1951kolmogorov}
F.~J. Massey~Jr, ``The {K}olmogorov-{S}mirnov test for goodness of fit,''
  \emph{Journal of the American statistical Association}, vol.~46, no. 253, pp.
  68--78, 1951.

\bibitem{wang2020detection}
C.~Wang, S.~Tindemans, K.~Pan, and P.~Palensky, ``Detection of false data
  injection attacks using the autoencoder approach,'' in \emph{2020
  International Conference on Probabilistic Methods Applied to Power Systems
  (PMAPS)}.\hskip 1em plus 0.5em minus 0.4em\relax IEEE, 2020, pp. 1--6.

\bibitem{energy_test}
G.~J. Sz{\'e}kely and M.~L. Rizzo, ``Energy statistics: A class of statistics
  based on distances,'' \emph{Journal of statistical planning and inference},
  vol. 143, no.~8, pp. 1249--1272, 2013.

\bibitem{code_energytest}
\BIBentryALTinterwordspacing
J.~Djolonga, ``A {P}y{T}orch library for differentiable two-sample tests,''
  2017. Accessed on: Oct. 3, 2021. [Online]. Available:
  \url{https://github.com/josipd/torch-two-sample/blob/master/docs/index.rst}
\BIBentrySTDinterwordspacing

\bibitem{liu2012high}
H.~Liu, F.~Han, M.~Yuan, J.~Lafferty, and L.~Wasserman, ``High-dimensional
  semiparametric {G}aussian copula graphical models,'' \emph{The Annals of
  Statistics}, vol.~40, no.~4, pp. 2293--2326, 2012.

\bibitem{goodfellow2014generative}
I.~Goodfellow, J.~Pouget-Abadie, M.~Mirza, B.~Xu, D.~Warde-Farley, S.~Ozair,
  A.~Courville, and Y.~Bengio, ``Generative adversarial nets,'' \emph{Advances
  in neural information processing systems}, vol.~27, 2014.

\bibitem{arjovsky2017wasserstein}
M.~Arjovsky, S.~Chintala, and L.~Bottou, ``Wasserstein generative adversarial
  networks,'' in \emph{International conference on machine learning}.\hskip 1em
  plus 0.5em minus 0.4em\relax PMLR, 2017, pp. 214--223.

\bibitem{ENTSO2020}
\BIBentryALTinterwordspacing
ENTSO-E, ``Mid-term adequacy forecast 2020,'' 2020. Accessed on: Oct. 1, 2021.
  [Online]. Available: \url{https://www.entsoe.eu/outlooks/midterm/}
\BIBentrySTDinterwordspacing

\bibitem{evans2020assessing}
M.~P. Evans and S.~H. Tindemans, ``Assessing energy storage requirements based
  on accepted risks,'' in \emph{2020 IEEE PES Innovative Smart Grid
  Technologies Europe (ISGT-Europe)}.\hskip 1em plus 0.5em minus 0.4em\relax
  IEEE, 2020, pp. 1109--1113.

\bibitem{pypi}
\BIBentryALTinterwordspacing
R.~T. McGibbon, ``quadprog 0.1.8 - pypi,'' 2021. Accessed on: Aug. 15, 2021.
  [Online]. Available: \url{https://pypi.org/project/quadprog}
\BIBentrySTDinterwordspacing

\end{thebibliography}

\end{document}